\documentclass[12pt,epsfig]{article}
\usepackage{amsmath}
\usepackage{graphicx}
\usepackage{epstopdf}
\usepackage{lscape}
\usepackage{longtable}
\usepackage[usenames, dvipsnames]{color}
\usepackage{caption}
\usepackage{subcaption} %for subfigures
\usepackage{amssymb}
\begin{document}
	\begin{center}
		\textbf {Confirmation of Binary Clustering in Gamma-Ray Bursts through an Integrated $p$-value from Multiple Nonparametric Tests of Hypotheses}\\
	\end{center}
	\begin{center}
		Dr. Soumita Modak\\
		Faculty of Statistics\\ 
		Basanti Devi College, University of Calcutta\\
		147B, Rash Behari Ave, Kolkata- 700029, India\\
		Email: soumitamodak2013@gmail.com\\
		Orcid id: 0000-0002-4919-143X\\
		Homepage: sites.google.com/view/soumitamodak
		
	\end{center}
	\begin{abstract}
		The paper applies a new, nonparametric, interpoint distance-based measure to confirm the inherent groups prevailing in the brightest source of light in the universe: gamma-ray bursts. Our effective metric, in association with clustering methods like Gaussian-mixture model-based and $K$-means algorithms, resolves the conflict regarding the possibility about existence of more than binary clusters in the gamma-ray burst population. Here we carry out multiple nonparametric statistical tests of hypotheses, as many as the number of bursts available from the `BATSE' catalog. An integrated $p$-value achieved from the aforesaid dependent tests solves our concern confirming two groups of short and long bursts.
	\end{abstract}
	\section{Introduction}
	Classification is a well-known statistical data mining technique broadly having two categories supervised and unsupervised. In the former, we are aware of the number of the prevailing groups and their properties, while the latter constitutes its unsupervised version, referred to as 
	`cluster analysis', usually with no information available regarding the clusters. In making the efforts to comprehend the origins causing the variation in an observed data set, we adopt clustering, where close members from the sample are clustered in the same group and far ones in different (Ripley 1996; Jain, Murty and Flynn 1999; Everitt, Landau and Leese 2001; Matioli et al. 2018; Cheng et al. 2021; Modak 2023a; Modak 2024c). It has its usefulness in numerous fields including astronomy to understand our ever expanding universe (Feigelson \& Babu 2013; Andreon \& Weaver 2015; Modak, Chattopadhyay \& Chattopadhyay 2017; Modak, Chattopadhyay \& Chattopadhyay 2020; Modak, Chattopadhyay \& Chattopadhyay 2022; Modak 2023d). 
	
	In the search of clusters in a new data set to uncover the potential sources (sometimes unprecedented) behind any existing grouping, it is extremely necessary to validate the answers of clustering algorithm, producing the clusters in the sample under investigation, to check the authenticity of the resulting clusters. Moreover, in order to remove the spurious outcome (if any) and to confirm the physical significance of the groups, it is important to subjectively explore the cluster-properties, in detail, using the domain-expertise. However, that is the final stage of cluster analysis, and succeeds the intermediate stage of cluster validation through the accuracy indices, that assess the quality of clusters obtained through the implemented clustering algorithm (Tibshirani, Walther \& Hastie 2001; Pakhiraa, Bandyopadhyay and Maulik 2004; Silva et al. 2020; Modak 2022; Modak 2023b; Modak 2023c; Modak 2023d; Modak 2024d; Modak 2024e). {In general, cluster analysis is extremely necessary machine learning method with the automatic algorithms for astronomical data sets, as the huge sample of observations is not feasible for the subjective classification of individual object (Chattopadhyay and  Chattopadhyay 2014; Modak 2023d). Especially, the ever evolving astrophysical phase of celestial bodies, might get into a stage either not known before or changed than earlier, can be explored only thorough clustering wherein no prior labels on them are available. On the other hand, clustering has the option to study properties of objects, transitioning from one stage to another, in terms of the share of characteristics from each of the stages (Modak 2021). Clustering is  valid when we reach the resulting clusters with physical sources. In contrast to supervised classification that forces the objects to get classified into previously established groups only, its unsupervised version in the form of clustering does not put any such stringent constraints and allows the data members for classification into some new cluster of an unprecedented source (if existing), which is extremely important to explore our universe for new discoveries (Modak 2019). In regard to astronomical data analysis, clustering approaches themselves are efficient enough or coupled with other machine learning techniques to deal with identifying noise, outlier, anomaly detection and all; where clustered sample is easy to interpret in terms of the cluster properties or prototypes rather than individual objects, which is very useful for data reduction and future classification guidance (Feigelson and
		Babu 2013; Modak et al. 2018; Modak et al. 2020).}
	
	The present study considers an important task of Astrostatistics, that concerns the clustering problem about the gamma-ray burst (GRB) population, where we analyze the data from the Burst and Transient Source Experiment (BATSE) catalog gathered through the COMPTON Gamma-Ray Observatory (CGRO) from year 1991 to 2000. Astrophysicists have so far confirmed two physically interpreted groups in GRBs (Kouveliotou et al. 1993; Woosley \& Bloom 2006; Nakar 2007; Berger 2014; Blanchard et al.
	2016). The groups typically respond to short (with $T_{90} < 2$ s) and long ($T_{90} > 2$ s) duration bursts, with $T_{90}$
	indicating the time of a burst's 90\% 
	flux arrival in seconds (s). By means of their spectral hardness ratios, it is found that the former cluster is hard and the latter is soft. They show significant difference in their prompt-emission, afterglows, host galaxies and redshift distributions (Kouveliotou et al. 1993; Gehrels et al. 2009; Zhang et al. 2012; Berger 2014; Levan et al. 2016). These distinct classes are believed to be resulted through compact binary mergers and massive stellar collapses respectively. 
	
	While huge diversity in GRBs led astronomers to their cluster analysis, a considerable overlap existing between the two physically interpreted groups with respect to their various properties made it necessary to look for additional clusters in the population (Tarnopolski 2016a; Chattopadhyay \& Maitra 2017). In this attempt, statistical analysis later found three possible groups (Mukherjee et al. 1998; Balastegui et al. 2001; Horv\'{a}th 2002; Chattopadhyay et al. 2007; King et al. 2007; Modak, Chattopadhyay \& Chattopadhyay 2018; Modak 2019), whereas recently five groups altogether emerged (Chattopadhyay \& Maitra 2017; Chattopadhyay \& Maitra 2018). The third group with an intermediate duration in terms of $T_{90}$ might be spurious occurred due to instrumental and
	sampling biases (Hakkila et al. 2000; Rajaniemi \& M\"{a}h\"{o}nen 2002; Hakkila et al. 2003; Tarnopolski 2019), whose physical source is yet to be ascertained; whereas the five-cluster scenario is dismissed as a result of apparently inappropriate parametric assumptions that were involved in the implemented clustering algorithms using Gaussian or $t$-mixture model-based methods (T\'{o}th, R\'{a}cz \& Horv\'{a}th 2019). T\'{o}th, R\'{a}cz \& Horv\'{a}th (2019) compared the cluster-properties and based on domain-expertise concluded the new two sub-groups as spurious partitions of the older three. As nonparametric methods are supposed to be more robust in such a situation (Bandyopadhyay and Modak 2018; Modak \& Bandyopadhyay 2019; Modak, Chattopadhyay \& Chattopadhyay 2020), we later on carry out a robust, nonparametric,
	fuzzy clustering method using FANNY algorithm (Kaufman and
	Rousseeuw 2005; Modak 2024a) on this data set. There we analyzed all the individual GRBs meticulously and drew statistically the same inference of discarding the two further groups (Modak 2021). There has also emerged, very recently, a possibility of four groups (Mehta and Iyyani 2024), based on the temporal and spectral properties
	using time-integrated spectra of prompt emission, through
	unsupervised nested Gaussian-mixture model classification in an unconventional manner, but the groups are found to possess division or mixture of the two known kinds of progenitors of the typical short and long bursts; whereas the opinions on the third group's spuriousness, although nothing confirmed yet physically, are changing (Modak, Chattopadhyay \& Chattopadhyay 2018; Tarnopolski 2022). 
	
	In the middle of this ongoing conflict regarding the actual number of groups with physical existence to explore the progenitors of this diverse grouping in the GRBs, the present paper implements a recently discovered method (Modak 2024b), effective enough for deciding on the number of clusters existing in a given set of data statistically; as only a correct statistical clustering would lead to the meaningful physical groups and not the spurious one(s). Our interpoint distance-based measure utilizes nonparametric statistical tests of hypotheses, as many as the number of bursts in the observed data set, in association with the widely used Gaussian-mixture model-based clustering (GMMBC; McLachlan and Peel 2000; Scrucca et al. 2016) and $K$-means algorithm (Hartigan 1975; Hartigan and Wong 1979), and confirms the two significant groups prevailing in the GRBs in terms of an integrated $p$-value achieved from the multiple tests. It is important to note that while a natural grouping of GRBs can be revealed by an appropriate clustering algorithm, the correct number of clusters can only be determined by a suitable cluster accuracy measure while implementing such classification methods like model or partitioning-based clustering that are not able to evaluate the number by themselves. This is the reason that model-based clustering methods led to a larger number of clusters in the past analyses (Chattopadhyay \& Maitra 2017; Chattopadhyay \& Maitra 2018; Mehta and Iyyani 2024). We solve this grave concern around these additional groups in this work. They are not because of the clustering methods in terms of parametric assumptions involved in the model-based clustering, rather due to the unsuitable accuracy measures implemented in association with them. For example, the Bayesian Information Criterion (BIC; Frayley and Raftery 1998) 
	indicated five clusters (Chattopadhyay \& Maitra 2017; Chattopadhyay \& Maitra 2018). The results of GMMBC with BIC using the same working variables on the same data set are reproduced by us, whereas another criterion: Integrated Complete-data Likelihood (ICL; Biernacki, Celeux and Govaert 2000) produced four clusters. On the other hand, our proposed integrated $p$-value criterion (Modak 2024b) resolves this conflict and confirms two statistically significant groups of the thus far recognized short and long bursts.
	
	The remainder of the paper is designed as follows: next section explains the data in detail, wherein the implemented clustering method is described in the section after the next. Section 4 does the clustering of GRBs and interprets the results, with the last section drawing the conclusion of the paper work.
	\section{Data}
	From the present BATSE Catalog\footnote{https://gammaray.nsstc.nasa.gov/batse/grb/catalog/current/}, we utilize values for the following variables observed for individual GRB.\\
	(i) $F_1$: time-integrated fluence in $20-50$ keV spectral channel,\\   
	(ii) $F_2$: time-integrated fluence in $50-100$ keV spectral channel,\\
	(iii) $F_3$: time-integrated fluence in $100-300$ keV spectral channel,\\
	(iv) $F_4$: time-integrated fluence in $>300$ keV spectral channel,\\      
	(v) $P_{256}$: peak flux measured in 256 ms bins,\\
	(vi) $T_{50}$: time within which $50\%$ of the flux arrives,\\
	(vii) $T_{90}$: time within which $90\%$ of the flux arrives.\\
	We also consider some experts' defined variables as follows.\\
	(viii) $F_T=F_1+F_2+F_3+F_4$: total fluence,\\
	(ix) $H_{32}=F_3/F_2$: spectral hardness ratio,\\
	(x) $H_{321}=F_3/(F_2+F_1)$: spectral hardness ratio.\\
	Unit of fluence: ergs per square centimeter (ergs cm$^{-2}$); unit of peak flux: count per square centimeter per second (cm$^{-2}$ s$^{-1}$); and unit of time: second (s). The present study involves a sample of size 1956 on the 6 working variables as: $log_{10}T_{50}, log_{10}T_{90}, log_{10}P_{256}, log_{10}F_{T}, log_{10}H_{32}$ and $log_{10}H_{321}$ (see, for details, Mukherjee et al. 1998; Hakkila et al. 2000; Horv\'{a}th et al. 2002; Chattopadhyay et al. 2007; Chattopadhyay and Maitra 2017; Toth et al. 2019; Modak 2021). 
	\section{Methodology}
	\subsection{Model-based clustering method}
	For given data set $D=\{M_1,...,M_n\}$, where $n$ is the sample size and $M_i$ denotes the $i$-th GRB member (equivalently, the corresponding observation vector) in the sample, we assume that a partition $P_K=\{C_1,...,C_K\}$, with the number of clusters = $K$, of the members exist, where $C_k$ is the $k$-th cluster of size $n_k$ such that $\sum_{k=1}^{K}n_k=n$. Then model-based clustering is constructed, assuming $M_i,i=1,2,...,n$ are independent and identically distributed through the probability density function of a finite mixture model of $K$ components in the following form (McLachlan and Peel 2000):
	\begin{equation}\label{MD}
		f(M_i,\boldsymbol{\theta})=\sum_{k=1}^{K}\pi_kf_k(M_i,\boldsymbol{\theta}_k),
	\end{equation}
	where $\boldsymbol{\theta}=(\pi_1,...,\pi_K,\boldsymbol{\theta}_1,...,\boldsymbol{\theta}_K)'$ is the vector of unknown parameters of the mixture model with $f_k(M_i,\boldsymbol{\theta})$ being the $k$-th component density for member $M_i$ with parameter vector $\boldsymbol{\theta}_k$, and $(\pi_1,...,\pi_K)'$ are the mixing weights satisfying  $\pi_k =Pr[M_i\in C_k]>0$ such that $\sum_{k=1}^{K}\pi_k=1$. For GMMBC, we take each $f_k$ to be a multivariate Gaussian density function with respective mean vector and dispersion matrix. $\boldsymbol{\theta}$ is estimated by the method of maximum likelihood with the help of the Expectation-Maximization algorithm (Scrucca et al. 2016).
	\subsection{Clustering accuracy measure}
	It is important to note that the model-based clustering is performed only with $K$ provided, which is unknown for clustering a real data set. Therefore, in practice, a proper cluster accuracy measure is suggested which can assess the quality of resulting clusters obtained through the implemented clustering algorithm. For model-based clustering, we have several specially designed model-based cluster accuracy measures available in the literature. For example, Bayesian Information Criterion (BIC; Frayley and Raftery 1998) and
	Integrated Complete-data Likelihood (ICL; Biernacki, Celeux and Govaert 2000), both involves maximum likelihood of model \eqref{MD} based on the given sample. The available approach to choose the number of clusters is to compute the values of the accuracy measures for each of $K=1,2,...,K_{\max}$ and finalize that with the optimal value of the corresponding measure, and the  respective value for $K$ is taken as the estimated value of unknown number of clusters present in the set of data. Here the maximum of the aforesaid measures, i.e. BIC and ICL, indicate the optimal value. Similar way is adopted for choosing a best model among many even with a fixed $K$. We use the inbuilt functions from `mclust' package of globally acknowledged statistical programming language `R' where diverse advanced Gaussian-mixture models are readily available, among which BIC and ICL indicate `VVV' to be the best, where the component densities are ellipsoidal with varying volume, shape, and orientation (see, Scrucca et al. 2016, for details). 
	
	Now, the biggest concern here is $K_{\max}$, obviously for clustering a data set with original $K$ unknown, it is crucial to choose this value. For example, if the chosen value is less than the real value of $K$, then it is impossible to reach the correct decision. Discussing with the current scenario of clustering GRBs, if we limited $K_{\max}$ to three, then the new possibilities regarding the additional clusters like four or five would have never been arrived. On the other hand, a larger value for $K_{\max}$ exceeding the original $K$ would only increase the unnecessary computational burden for this large data set, for example we consider, in the present study, $K_{\max}=10$ as we want to carry out the clustering of the bursts afresh. Therefore, we propose a new accuracy measure (Modak 2024b), which is not only applicable to results beyond model-based clustering but also solves the above-mentioned major issues. It starts considering from $K=1$ and increases its value by one post-rejection of the previous value while carrying on the clustering simultaneously, and thus stops at the estimated $K$. This measure has been evidenced to perform excellent in finding the correct number of clusters in a data set. Therefore, we use it in association with the GMMBC (with model VVV), whose detailed description is provided in the next section (Modak 2024b).
	\subsection{Clustering algorithm}
	(a) We adopt the popular GMMBC, which is applied to the above data for $K=2$ as a priori.\\ 
	(b) Now from the resulting binary classification, we select a GRB (say the $i$-th) and compute its interpoint distances from all other GRBs in the data set. Thus we obtain two populations of distances: (i) the cluster to which the $i$-th GRB is clustered through step (a), and (ii) the other cluster.\\
	(c) All interpoint distances from (b) are normalized to belong to the semi-open interval $(0, 1]$, which is then divided into $(w-1)$ number of equally-spaced, mutually
	exclusive and exhaustive sub-intervals. It makes $(w-1)$ similarly constructed categories for both populations, where we take into account the frequencies of distances falling in each of the sub-intervals for the two populations separately.\\
	(d) Next, we carry out the well-known nonparametric homogeneity test for testing the hypothesis of equality for the populations in terms of the interpoint distances from 2 clusters (using the data from c), where the test statistic follows an approximate chi-square distribution, with degree(s)
	of freedom $=\overline{w-1}-1=w-2$ (Hogg, Mckean, and Craig 2019).%pg 303, s/c). 
	
	Subsequently, we find the corresponding $p$-value as $p_i$ for the $i$-th GRB
	with
	\begin{equation}
		p_i=Prob_{H_0}\hspace{.05in}[T_i>t_i],  
	\end{equation}
	where $T_i$ is the test statistic following approximately $\chi^2_{(w-2)}$ for testing $H_0:K=1$ vs $H_1:K>1$, with $t_i$ as its observed value for the sample while considering $i$-th burst. The null is rejected if $p_i<\alpha$, level of significance for the test of hypothesis; otherwise, the alternative is rejected. The interpretation of these hypotheses in the context of evaluating $K$ stands as: the null says that the two populations are almost the same in which the $i$-th GRB is clustered and not clustered, consequently, based on the $i$-th GRB, it is concluded that $K=1$; whereas in case of the alternative, at least two clusters are accepted to be possible.\\
	(e) In order to explore the clustering regarding the complete data set by looking at one individual GRB at a time, we redo everything from (b) to (d) for each burst, i.e. $i=1(1)1956$, which constitute our multiple tests of hypotheses (Jure\v{c}kov\'{a} and Kalina 2012; Modak \& Bandyopadhyay 2019) based on the vector of test statistics: $(T_1,...,T_{1956})'$ leading to the vector of required $p$-values $(p_1,....,p_{1956})'$.\\
	(f) To reach a decision at this stage, we require an integrated $p$-value (Kost and McDermott 2002; Poole et al. 2016; Modak \& Bandyopadhyay 2019), which shall make the conclusion for the entire data set of all GRBs. Ours denoted by `IP' is defined, based on all dependent $p$-values obtained from (e), as follows:
	\begin{equation}
		\text{IP}=E(p^*)=\sum_{i=1}^{N}p_i/N, 
	\end{equation}
	where \begin{equation*}
		Prob\hspace{.05in}[p^*=p_i]=1/N,\forall i=1,2,...,N,
	\end{equation*}
	and
	\begin{equation*}
		Prob\hspace{.05in}[p^*<\alpha]=\sum_{i=1}^{N}Prob\hspace{.05in}[p_i<\alpha]/N,
	\end{equation*} 
	$\alpha$ is the common level of significance selected for all the individual tests.
	
	Two possible outcomes are reached regarding the clustering of the given sample in terms of our measure: either (i) IP $\geq \alpha$ with acceptance of the null hypotheses of equality for the two populations under investigation and we conclude that there is just $K=1$ cluster in the data as a whole, or (ii) IP $< \alpha$ with rejection of the aforesaid nulls and we draw the inference that there are at least two clusters prevailing in the GRB data set, following which we proceed to look for more clusters in the next level.  \\
	(g) 3-component GMMBC is performed on the sample.\\ 
	(h) Analogous to (b), for the $i$-th GRB, we compute the corresponding two populations of interpoint distances. Here one population corresponds to the cluster in which the $i$-th GRB is clustered through stage (g), whereas the  other population is that cluster which is nearest to the $i$-th GRB (i.e., with a lower mean distance from all members of the cluster) between the other two resulting clusters from (g).\\
	(i) Repeat steps (c) through (f), where either exactly two number of clusters is accepted, or `at least $K=3$' is confirmed followed by further search for a larger number in a similar manner, which continues until we find a definite value for $K$.
	
	We also perform $K$-means method (Hartigan 1975) using the Hartigan-Wong algorithm (Hartigan and Wong 1979), known to produce good clustering in GRBs (Chattopadhyay et al. 2007; Modak, Chattopadhyay \& Chattopadhyay 2018; Modak 2019), in the place of GMMBC on our data using the above algorithm.
	\section{Data analysis}
	First, we report the answers of GMMBC through BIC and ICL, as widely used,
	that respectively give rise to $\hat{K}=5$ and 4 (see, Table~\ref{BIC-ICL}). These results are in accordance with the findings by Chattopadhyay \& Maitra (2017) and Chattopadhyay \& Maitra (2018); and Mehta and Iyyani (2024) respectively. Now, we verify the consequences of GMMBC through our suggested method from Section 3.3.
	
	While implementing our new technique for determination of $K$, we have two tuning parameters involved in it: $w$, inherently attached to our measure IP and $\alpha$ associated with our approach.
	Extensive data study from Modak (2024b) leads us to the choice of $w=3$ and $\alpha=0.01$, with the minimum computation and accurate results. Here our approach says there are exactly two clusters prevailing in the GRBs (Table~\ref{pvalueGMMBC}).
	
	Prominently, this significantly varying answers, found in the last two paragraphs, reveal the difficulty in clustering the GRB data set. Therefore, the distinct statistical methods applied to various variables from the past literature could not solve the global issue around the possible number of clusters in existence. However, we do it by quantifying the quality of the clustering output generated by distinct measures with utilizing multiple popular cluster accuracy indices, described below.
	
	We are to assess a partition $P_K=\{C_1,...,C_K\}$, where $C_k$ is the $k$-th cluster of size $n_k$ such that $\sum_{k=1}^{K}n_k=n$, and $M_{km}$ is the $m$-th member clustered in $C_k$ for $m=1,2,...,n_k,k=1,2,...,K$. It is to be noted that for a fixed value of $K$, $P_K$ is not unique as the clusters can have different numbers of members or the same numbers with different members. Among all those partitions with differing values of $K$ as well, we need to find out the best possible one so that heterogeneity between the different clusters and homogeneity within the clusters are as high as possible.
	
	(i) The average silhouette width (ASW; Rousseeuw 1987):
	\begin{equation}
		\text{ASW}=\frac{1}{n}\sum\limits_{k=1}^{K}\sum\limits_{m=1}^{n_k}s(M_{km}),
	\end{equation}
	where the silhouette width of the member $M_{km}$:
	\begin{equation*}
		s(M_{km})=\frac{b(M_{km})-a(M_{km})}{\max\bigg\{a(M_{km}),b(M_{km})\bigg\}},
	\end{equation*}
	with
	\begin{equation*}
		a(M_{km})=\sum\limits_{m'=1}^{n_{k}}d(M_{km},M_{km'})\big{/}(n_{k}-1)
	\end{equation*}
	and	
	\begin{equation*}
		b(M_{km})=\underset{1\leq k'(\neq k)\leq K}{\min}\bigg\{\sum\limits_{m'=1}^{n_{k'}}d(M_{km},M_{k'm'})\big{/}n_{k'}\bigg\}
	\end{equation*}.
	ASW $\in [-1,1]$: max value is coveted.

	(ii) The Dunn index (Dunn; Dunn 1974):
	\begin{equation}
		\text{Dunn}=\frac{\underset{k\neq k'}{\underset{1\leq k,k' \leq K}{\min}}\bigg\{ \underset{ 1\leq m \leq n_k, 1\leq m' \leq n_{k'}}\min d(M_{km},M_{k'm'})\bigg\}}{\underset{1\leq k \leq K}\max \bigg\{\underset{1\leq m,m' \leq n_k}\max d(M_{km},M_{km'})\bigg\}},
	\end{equation}
	Dunn$\in(0,\infty)$: highest value is desired.\\
	
	(iii) Cali\'{n}ski and Harabasz index (CH; Cali\'{n}ski \& Harabasz 1974):
	\begin{equation}
		\text{CH}=\frac{\sum\limits_{k=1}^{K}n_k d(\overline{M}_{k0}\large{,}\overline{M}_{00})/(K-1)}{\sum\limits_{k=1}^{K}\sum\limits_{m=1}^{n_k}d(M_{mk}\large{,}\overline{M}_{k0})/(n-K)}, 
	\end{equation}
	where
	$\overline{M}_{k0}=\frac{1}{n_k}\sum\limits_{m=1}^{n_k}M_{km}$ and $\overline{M}_{00}=\frac{1}{n}\sum\limits_{k=1}^{K}n_k\overline{M}_{k0}$. CH$\in(0,\infty)$ is maximized for the best classification
	
	(iv) Connectivity (Conn; Handl et al. 2005):
	\begin{equation}
		\text{Conn}=\sum\limits_{k=1}^{K}\sum\limits_{m=1}^{n_k}\sum\limits_{j=1}^{J}I_{km}(j), 
	\end{equation}
	where $I_{km}(j)=0$ with the $j$-th nearest member of $M_{km}\in C_k$, otherwise $I_{k,m}(j)=1/j$. We choose $J=10$ for the present study. Conn$\in[0,\infty)$: lower value indicates better clustering.
	
	(v) Nearest neighbor classification error rate:
	(NNCER; Ripley 1996):\\
	Here we consider a statistic $I_l(k,m)=0$ if majority of the $l$ nearest neighbors (NNs) of member $M_{km}$ are also clustered in $C_k$, otherwise  $I_l(k,m)=1$; whereas tie by $l$ NNs is broken through choosing $I_l(k,m)=0$ or $1$ randomly. Then the targeted measure is defined as 
	\begin{equation}
		\text{NNCER}=\frac{1}{n}\sum\limits_{k=1}^{K}\sum\limits_{m=1}^{n_k}I_l(k,m), 
	\end{equation}
	with $l=9$ for our study. NNCER$\in[0,1]$ is minimized for the optimal clustering.
	
\begin{figure}
		\centering
		\includegraphics[width=1\textwidth]{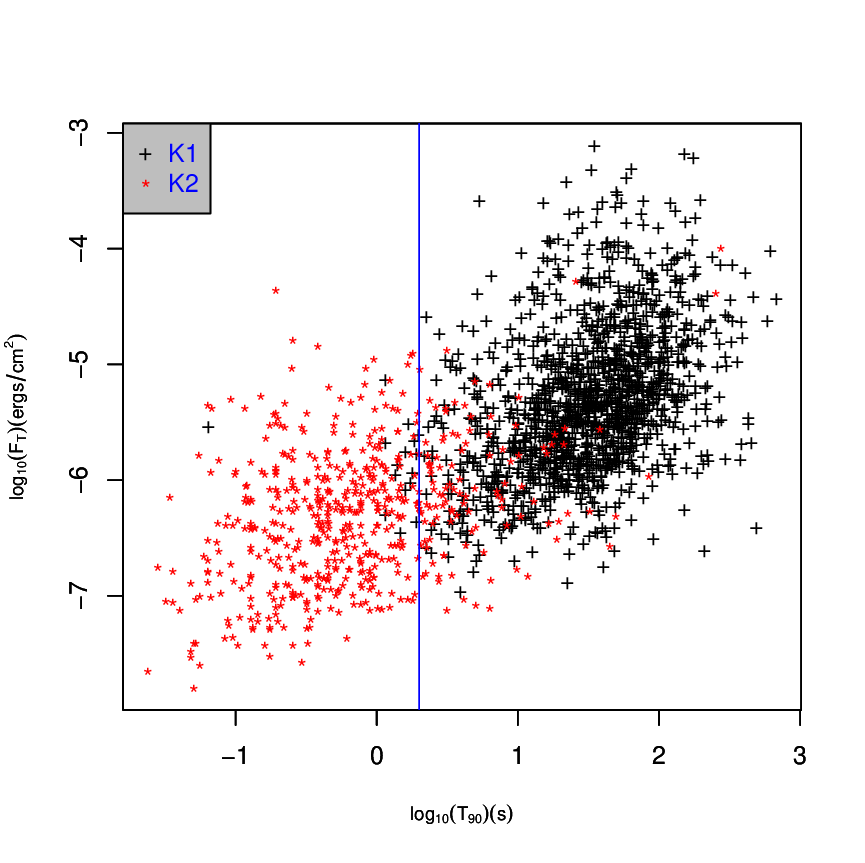}
		\caption{Plot of $log_{10}(T_{90})$ (in s) vs. $log_{10}(F_{T})$ (in ergs cm$^{-2}$) for two clusters of GRBs from GMMBC, wherein the vertical blue line represents $T_{90}=2$ s.}\label{f1_GMMBC}
	\end{figure}

\begin{figure}
		\centering
		\includegraphics[width=1\textwidth]{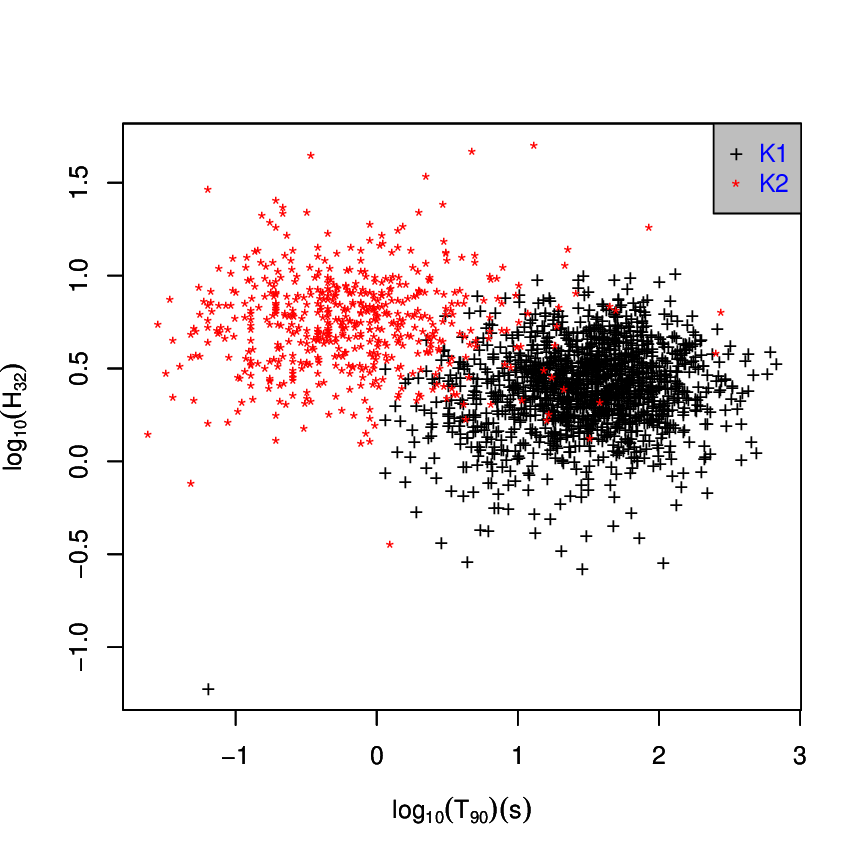}
		\caption{Plot of $log_{10}(T_{90})$ (in s) vs. $log_{10}(H_{32})$ for two clusters of GRBs from GMMBC.}\label{f2_GMMBC}
	\end{figure}

\begin{figure}
		\centering
		\includegraphics[width=1\textwidth]{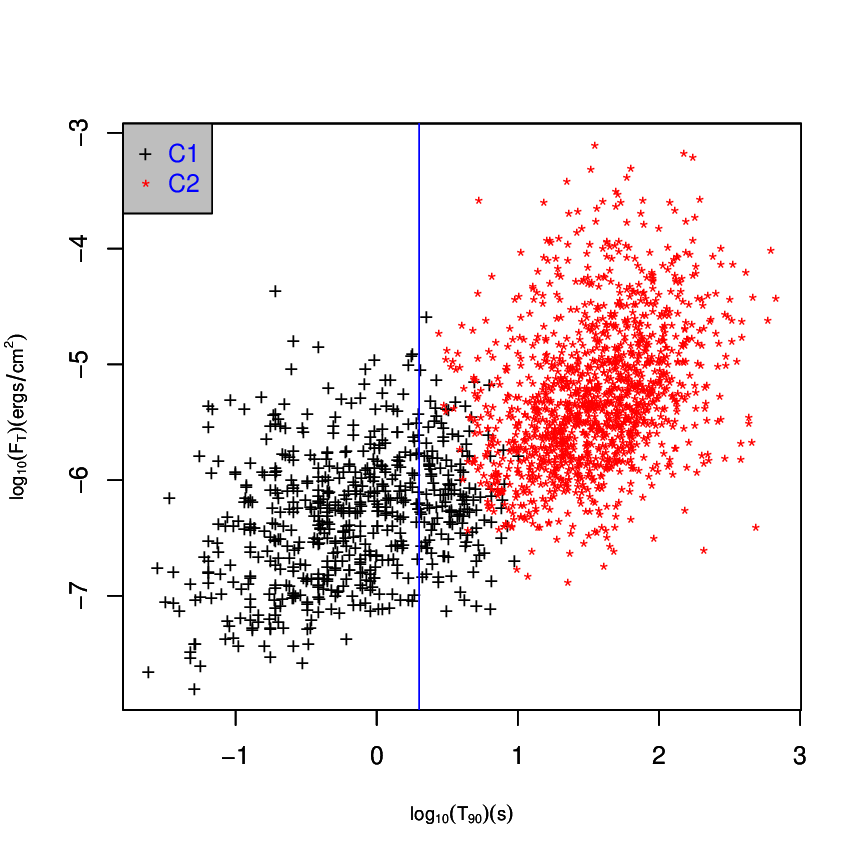}
		\caption{Plot of $log_{10}(T_{90})$ (in s) vs. $log_{10}(F_{T})$ (in ergs cm$^{-2}$) for two clusters of GRBs from $K$-means clustering, wherein the vertical blue line represents $T_{90}=2$ s.}\label{f1_Kmeans}
	\end{figure}

\begin{figure}
		\centering
		\includegraphics[width=1\textwidth]{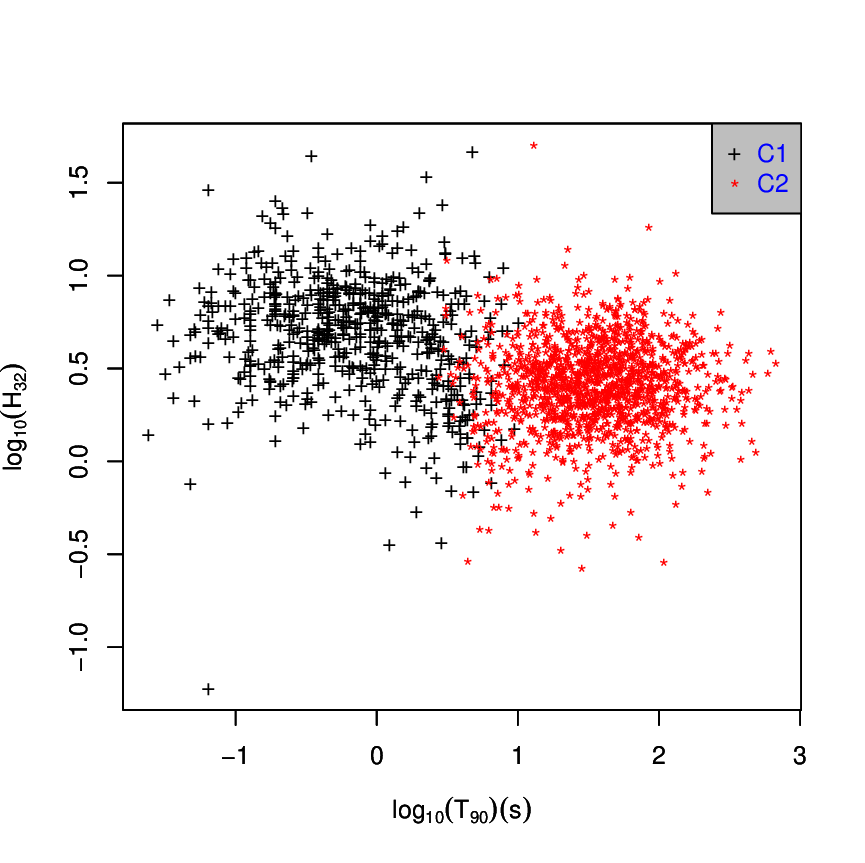}
		\caption{Plot of $log_{10}(T_{90})$ (in s) vs. $log_{10}(H_{32})$ for two clusters of GRBs from $K$-means clustering.}\label{f2_Kmeans}
	\end{figure}

	Comparison of the three different answers, found from the same clustering method GMMBC but by independent accuracy metrics (namely, BIC, ICL and IP), are shown in Table~\ref{CI}. It certainly manifests two significant clusters statistically, in terms of all the distinct assessment indices consistently and significantly, whose properties are displayed in Table~\ref{CPGMMBC} and Figures~\ref{f1_GMMBC}-\ref{f2_GMMBC}. Next, we study the impact of $K$-means clustering method in association with our measure IP on the current data, which also suggests two clusters (Table~\ref{pvalue_Kmeans}). The corresponding cluster-properties are reported in Table~\ref{CPKmeans} and Figures~\ref{f1_Kmeans}-\ref{f2_Kmeans}.
	
	Finally, comparison of GMMBC results with those of $K$-means prominently confirms the binary grouping pattern through the correct $K$-selector: our newly discovered integrated $p$-value (i.e. IP; Modak 2024b). The Tables and figures included in the paper state that groups K2 and K1 resulted from GMMBC are respectively comparable with the clusters C1 and C2 from $K$-means; whereas their drawn figures (Figures~\ref{f1_GMMBC}-\ref{f2_Kmeans}) comprehensibly exhibit two overlapping classes through both the clustering algorithms. However, $K$-means is undeniably superior as indices: (ASW, Dunn, CH, Conn, NNCER) = $(0.494,0.037,2476.8,100.313,0.767)'$ for the two clusters by it, are better than those of GMMBC (see, second column of Table~\ref{CI}). Moreover, the group duo in GMMBC have some of the GRBs clearly misclassified, because they show properties of the other group and hence are outliers in the group they are clustered to (see, Figures~\ref{f1_GMMBC}-\ref{f2_GMMBC}).
	
	Therefore, the variables are further interpreted with respect to the statistically found two clusters from $K$-means procedure.
	From Table~\ref{CPKmeans} and Figures~\ref{f1_Kmeans}-\ref{f2_Kmeans}, we see that cluster C1 consists of 31.2\% of the bursts having short duration with low total fluence (in terms of the average $F_T$) and lower peak flux (i.e. $P_{256}$) indicating fainter bursts, whose average $T_{90}$ is less than 2 s as per the conventional classification (Horv\'{a}th 2002); however, individual short bursts are exceeding this limit (Fig.~\ref{f1_Kmeans}; Tarnopolski 2015). The other group C2 has long and more luminous bursts (with higher $F_T$ and $P_{256}$), which are softer compared to the short ones (measured by $H_{32}$ and $H_{321}$, where lower values indicate hard burst). {It is important to note that the present analysis is data-driven, and therefore may alter with a change in the data in terms of the working variables and/or other GRB catalog. In this context, our findings are consistent with previously reached two groups from other earlier works on GRBs: univariate clustering on duration of short and long, bivariate classification of time and spectral-hardness-ratio plane of short-hard and long-soft, or multivariate pattern recognition of short-hard-faint and long-soft-bright (Kouveliotou et al. 1993; Woosley \& Bloom 2006; Nakar 2007; Berger 2014; Blanchard et al. 2016; Tarnopolski 2015; 2019). Moreover, it is relevant to draw a must-do comparison that our resulting two-cluster scenario is also consistent with the findings by analyses of bursts from other observed data sets: e.g. Venera GRB data gathered by the KONUS experiment (Mazets et al. 1981;
		Norris et al. 1984); bursts from the GRANAT experiment (Dezalay et al. 1992); RHESSI GRBs (\v{R}\'{\i}pa et al. 2012); Suzaku sample (Ohmori et al. 2016, Tarnopolski 2022), Fermi data set (Narayana Bhat et al. 2016; Kulkarni and Desai 2017; Zitouni et al. 2018), Swift bursts (Tarnopolski 2016b; Yang et al. 2016), the GRB Big Table and Greiner’s GRB catalog (Luo et al. 2023), and so on}. 
	
	{In regard to physical characteristics leading to the possible progenitors behind these two groups, we can comment: bursts from the cluster C1 are believed to have compact binary mergers, like merger of two neutron stars or merger
		of a neutron star with a black hole, as their progenitors, and the long ones from C2 with higher brightness have massive stellar collapse (Paczyński 1986; Usov 1992; Nakar 2007; Berger 2011; Zhang et al. 2012; Berger 2014; Blanchard et al. 2016; Modak et al. 2018; Tarnopolski 2019), based on their
		physical differentiation in terms of the prompt emissions, afterglows, host galaxies, redshift distributions
		(Kouveliotou et al. 1993; Gehrels et al. 2009; Berger 2011; Zhang et al. 2012; Levan et al. 2016); or other observational properties like supernova, kilonova, and,
		gravitation wave (Troja et al. 2019; Minaev \& Pozanenko 2020; Zhu et al. 2024). As a GRB associated with a supernova is known to have a core collapsar origin (Melandri et al. 2019; Minaev \& Pozanenko 2020), and an association with a kilonova (Lamb et al. 2019; Troja et al. 2019) indicates to have been originated from the merger of binary neutron
		stars (Goldstein et al. 2017; Wang et al. 2017).}
	\section{Conclusion}
	In this study, we apply a new statistical technique to find out the number of existing clusters in GRBs while solving the conflict whether more than binary grouping is prevailing or not. Our approach re-verified the existence of two groups of short and long bursts, consistent with the so far known kinds of sources, whose further property-study is to be done on the astrophysical ground. Note that the present work revolves around the grouping based on statistical metrics; and the two resulting groups, overlapping in their properties, need additional investigation: with more precise future observations and improved physical models, to confirm their progenitors generally referred to as non-collapsar and collapsar respectively behind the short and long GRBs. We have statistically shown through automatic clustering algorithm, based on the present BATSE catalog under a multivariate set-up, that natural two groups in GRBs can be retrieved by GMMBC or $K$-means clustering method in terms of an appropriate accuracy index like ours; whereas the additional groups greater than the binary classes are concluded to be spurious in the sense that they were numerically found due to implementation of other inappropriate accuracy measures, namely BIC and ICL, but have no physical interpretation in terms of astrophysical facts known till date.
	
	{We conclude the work highlighting the wide applicability of our proposed method in determining the unknown number of clusters for any astronomical data sets in general. It is compatible with arbitrary clustering algorithms, needed $K$ as a priori, e.g. it can be used with nonparametric algorithms like $K$-means; or model-based clusterings like GMMBC, however, not restricted to them such as model-based indices BIC or ICL. Its efficacy is extensively explored with respect to data analysis in Modak (2024b); which is applicable to any-dimensional space, such as to rare astronomical sample where the number of objects could be close to or even less than the number of study variables, due to the measure's intrinsic design based on the interpoint distances. Moreover, the distances make it usable with any kind of data: continuous, discrete, or observations not even measurable but available in other forms of attributes or mixed of them with variables (Modak 2024a). Unlike most other measures, it does not need to be computed for all values of $K$ provided such as $K=2,3,4...$, which involves the tough decision on where to stop. Rather our approach performs the evaluation in a step-wise process ans stops at a finite step with $K$ estimated by reducing computational burden, which is very much needed for big celestial data sets. Indices following the former approach (like BIC or ICL) often posses very close values over different $K$ for challenging data sets, and hereby $K$ estimation could go wrong (Chattopadhyay \& Maitra 2017). Also for a new data set, our novel index can check whether there is any clustering pattern at all by confirming $K=1$ against $K$ is greater than one initially, and then only proceeds with further analysis. Therefore, it is a very appealing measure proposed for astronomical objects' clustering, and analysts are encouraged to carry on with its future application to other GRB catalogs in order to confirm the value of $K$ for this population.}
	\clearpage
	\begin{table}
		\caption{Values of model-based cluster accuracy indices namely, BIC and ICL, while clustering GRBs along with GMMBC for different number of clusters ($K$) as a priori (optimal values of the indices are marked in bold).}
		\begin{center}
			\begin{tabular}{c|c|c}
				\hline
				$K$&BIC&ICL\\\hline
				1&-5604.961&-5604.961\\[1ex]
				2&-4292.088&-4389.485\\[1ex]
				3&-3532.902&-4068.259\\[1ex]
				4&-3136.202&\textbf{-3813.228}\\[1ex]
				5&\textbf{-3040.254}&-3861.700\\[1ex]
				6&-3190.559&-4240.143\\[1ex]
				7&-3278.440&-4338.399\\[1ex]
				8&-3151.255&-4169.121\\[1ex]
				9&-3245.672&-4308.250\\[1ex]
				10&-3306.636&-4462.997\\[1ex]
				\hline
			\end{tabular}
		\end{center}
		\label{BIC-ICL}
	\end{table}
	\clearpage
	\begin{table}
		\caption{Values of our proposed integrated $p$-value at different stages while clustering GRBs through GMMBC and the corresponding decisions.}
		\begin{center}
			\begin{tabular}{cccc}
				\hline\\
				IP$_{w=3,\alpha=0.01}$&Null&Alternative& Decision\\	
				&hypothesis&hypothesis&\\
				\hline\\
				0.00397&$K=1$&$K>1$&Reject\\[1ex]
				0.0617&$K=2$&$K>2$&Accept\\[1ex]\hline
			\end{tabular}
		\end{center}
		\label{pvalueGMMBC}
	\end{table}
	\clearpage
	\begin{table}
		\caption{Values of multiple cluster accuracy indices for clustering GRBs through GMMBC with different number of clusters ($K$) as a priori (optimal values of the indices are marked in bold).}
		\begin{center}
			\begin{tabular}{c|c|c|c}
				\hline
				Index &$K=2$&$K=4$&$K=5$\\	\hline
				ASW& \textbf{0.473}&0.162&0.085\\[1ex]
				Dunn&\textbf{0.021}&0.015&0.013\\[1ex]		
				CH&\textbf{2136.461}&734.481&528.972\\[1ex]
				Conn&\textbf{149.291}&1065.417&1488.78\\[1ex]
				NNCER&\textbf{1.994}&15.491&21.37\\[1ex]\hline
			\end{tabular}
		\end{center}
		\label{CI}
	\end{table}
	\clearpage
	\begin{table}
		\caption{Properties of two groups of GRBs (mean value with standard error) through GMMBC.}
		\begin{center}
			\tiny
			\begin{tabular}{c c c c c c c c}
				\hline\\
				Cluster&Cluster-size & $T_{50}$ &  $T_{90}$ & $P_{256}$ & $F_{T} \times 10^{6}$ & $H_{32}$ &$H_{321} $\\
				name &  (percentage)        & (s)     & (s)     & (cm$^{-2}$ s$^{-1}$) &  (ergs cm$^{-2}$)      & &\\[1ex]
				\hline\\
				K1       & 1391 (71.1\%)  & 22.759 $\pm$ 1.090& 52.905$\pm$1.762&  3.620$\pm$0.238&  17.224$\pm$1.358&
				2.814$\pm$0.040&  1.629$\pm$ 0.027\\[1ex]
				
				K2       & 565 (28.9\%)  &  0.746 $\pm$ 0.131 & 2.969$\pm$0.701 & 2.309$\pm$0.158& 1.519$\pm$0.233& 6.621$\pm$ 0.210 & 3.960$\pm$0.095\\[1ex]
				
				\hline
			\end{tabular}
		\end{center}
		\label{CPGMMBC}
	\end{table}
	\clearpage
	\begin{table}
		\caption{Values of our proposed integrated $p$-value at different stages while clustering GRBs through $K$-means algorithm and the corresponding decisions.}
		\begin{center}
			\begin{tabular}{cccc}
				\hline\\
				IP$_{w=3,\alpha=0.01}$&Null&Alternative& Decision\\	
				&hypothesis&hypothesis&\\
				\hline\\
				0.00472&$K=1$&$K>1$&Reject\\[1ex]
				0.02921&$K=2$&$K>2$&Accept\\[1ex]\hline
			\end{tabular}
		\end{center}
		\label{pvalue_Kmeans}
	\end{table}
	\clearpage
	\begin{table}
		\caption{Properties of two groups of GRBs (mean value with standard error) through $K$-means clustering.}
		\begin{center}
			\tiny
			\begin{tabular}{c c c c c c c c}
				\hline\\
				Cluster&Cluster-size & $T_{50}$ &  $T_{90}$ & $P_{256}$ & $F_{T} \times 10^{6}$ & $H_{32}$ &$H_{321} $\\
				name &  (percentage)        & (s)     & (s)     & (cm$^{-2}$ s$^{-1}$) &  (ergs cm$^{-2}$)      & &\\[1ex]
				\hline\\
				C1       & 610 (31.2\%)  & 0.489 $\pm$ 0.023  & 1.365 $\pm$ 0.066 & 2.272 $\pm$ 0.143
				&1.2 $\pm$  0.107 & 5.975 $\pm$ 0.188 &  3.637 $\pm$ 0.095\\[1ex]
				
				C2       & 1346 (68.8\%)  &  23.611 $\pm$ 1.120 & 55.301 $\pm$ 1.811& 3.681 $\pm$ 0.246 
				&17.894 $\pm$ 1.403 & 2.980 $\pm$ 0.057 & 1.698 $\pm$ 0.030\\[1ex]
				
				\hline
			\end{tabular}
		\end{center}
		\label{CPKmeans}
	\end{table}
	\clearpage
	\clearpage
	\clearpage
	\clearpage
	\section*{Data Availability Statement}
	The source of the real data used for the present work is mentioned in the manuscript.
	\section*{Disclosure statement}
	No potential conflict of interest was reported by the author.
	\section*{Funding}
	No funding received for this work.
%	\section*{Acknowledgments}
%	The present work on Astrostatistics is dedicated to my beloved Doctoral (Joint) Supervisor: Late Prof. Dr. Mrs. Tanuka Chattopadhyay from the Department of Applied Mathematics, Former Dean of Science, University of Calcutta, Kolkata, India. 
%	I would also like to take this opportunity to thank my Doctoral Supervisor, Teacher, Mentor and Academic Mother: Prof. Dr. Mr. Asis Kumar Chattopadhyay from the Department of Statistics, Former Vice-Chancellor of University of Calcutta, Kolkata, India,
%	who encouraged me to work on GRBs. {The author expresses her sincere gratitude to the Editors for guiding through the entire process and conveying their comments elaborately with ease. Author thanks two anonymous reviewers for their expert advice and providing constructive comments to revise the manuscript in order to make it more useful for readers.}
%	\clearpage
	{}

\begin{thebibliography}{}
		\bibitem{}
		Andreon, S. \& Weaver, B. (2015), Bayesian Methods for the Physical
		Sciences– Learning from Examples in Astronomy and Physics, Springer
		Series in Astrostatistics, 4, Springer International Publishing, Switzerland.
		\bibitem{}
		Balastegui, A., Ruiz-Lapuente, P., \& Canal, R. (2001). \textsl{Reclassification of gamma-ray bursts.} Monthly Notices of the Royal Astronomical Society. \textbf{328}, 283--290.
		\bibitem{}
		Bandyopadhyay, U. and Modak, S. (2018). \textsl{Bivariate density estimation using normal-gamma kernel with application to astronomy}, Journal of Applied Probability and Statistics, \textbf{13}, 23-39. 
		\bibitem{}	
		Biernacki, C., Celeux, G., and Govaert, G. (2000). \textsl{Assessing a mixture model for clustering with the integrated completed likelihood}. IEEE Trans. Pattern Analysis and Machine Intelligence, \textbf{22}, 719-725.
		\bibitem{}
		Blanchard, P. K., Berger, E., Fong, W.--f. (2016). \textsl{The Offset and Host Light Distributions of Long Gamma-Ray Bursts: A New View from HST Observations of Swift Bursts}. The Astrophysical Journal. \textbf{817}, 144.
		\bibitem{}
		Berger, E. (2011). \textsl{The environments of short-duration gamma-ray bursts and implications for their progenitors.} New Astronomy Reviews. \textbf{55}, 1--22.
		\bibitem{}
		Berger, E. (2014). \textsl{Short-Duration Gamma-Ray Bursts.} Annual Review of Astronomy and Astrophysics. \textbf{52}, 43--105.
		\bibitem{}
		Blanchard, P. K., Berger, E., Fong, W.--f. (2016). \textsl{The Offset and Host Light Distributions of Long Gamma-Ray Bursts: A New View from HST Observations of Swift Bursts}. The Astrophysical Journal. \textbf{817}, 144.
		\bibitem{}
		Cali\'{n}ski, T. \&  Harabasz, J. (1974). \textsl{A Dendrite Method for Cluster Analysis}. Communications in Statistics -- Theory and Methods. \textbf{3}, 1--27.
		{
			\bibitem{}
			Chattopadhyay, A. K., and  Chattopadhyay, T. (2014). \textsl{Statistical
				methods for astronomical data analysis}. Springer: New York.}
		\bibitem{}
		Chattopadhyay, S. \& Maitra, R. (2017). \textsl{Gaussian-mixture-model-based cluster analysis finds five kinds of gamma-ray bursts in the BATSE catalogue}. Monthly Notices of the Royal Astronomical Society. \textbf{469}, 3374--3389.
		\bibitem{}
		Chattopadhyay, S. \& Maitra, R. (2018). \textsl{Multivariate 
			$t$-mixture-model-based cluster analysis of BATSE catalogue
			establishes importance of all observed parameters, confirms five distinct
			ellipsoidal sub-populations of gamma-ray bursts}. Monthly Notices of the Royal Astronomical Society. \textbf{481}, 3196--3209.
		\bibitem{}
		Chattopadhyay, T., Misra, R., Chattopadhyay, A. K., \& Naskar, M. (2007). \textsl{Statistical evidence for three classes of gamma-ray bursts}. The Astrophysical Journal. \textbf{667}, 1017--1023.
		\bibitem{}
		Cheng, D., Zhu,  Q., Huang, J., Wu, Q. and Yang, L. (2021).  \textsl{Clustering with Local Density Peaks-Based Minimum Spanning Tree}. IEEE Transactions on Knowledge and Data Engineering. \textbf{33}, 374--387.
		{
			\bibitem{}
			Dezalay J.-P., Barat C., Talon R., Syunyaev R., Terekhov O., Kuznetsov A.:
			(1992), in Paciesas W. S., Fishman G. J., eds, AIP Conf. Ser. Vol. 265,
			Huntsville GRB Workshop. Am. Inst. Phys., New York, Page: 304.}
		\bibitem{}
		Dunn, J. C. (1974). \textsl{Well-separated clusters and optimal fuzzy partitions.}  Journal of Cybernetics. \textbf{4}, 95--104.
		\bibitem{}
		Everitt, B. S., Landau, S. and Leese,  M. (2001). \textsl{Cluster Analysis.} Arnold, London.
		\bibitem{}
		Frayley, C. and A. E., Raftery (1998). \textsl{How many clusters? Which clustering method? Answers via model based
			cluster analysis.} The Computer Journal, \textbf{41}, 578–88.
		\bibitem{}
		Feigelson, E. D. \& Babu, G. J. (Eds.) (2013), Statistical Challenges in
		Modern Astronomy V, Lecture Notes in Statistics - Proceedings, 209,
		Springer Science+Business Media, New York.
		{
			\bibitem{}
			Goldstein, A., Veres, P., Burns, E., Briggs, M. S., Hamburg,  R., Kocevski, D., Wilson-Hodge, C. A., Preece, R. D., Poolakkil, S.,  Roberts, O. J., Hui, C. M., Connaughton, V., Racusin, J., von Kienlin,  A., Canton, T. D., Christensen, N., Littenberg, T., Siellez,  K., Blackburn, L., Broida, J., Bissaldi, E., Cleveland, W. H., Gibby, M. H., Giles, M. M., Kippen, R. M. , McBreen, S., McEnery, J., Meegan, C. A., Paciesas, W. S. and Stanbro, M. (2017). \textsl{An Ordinary Short Gamma-Ray Burst with Extraordinary Implications: Fermi-GBM Detection of GRB 170817A}. The Astrophysical Journal Letters, \textbf{848}, Article id: L14, 14 Pages.}
		\bibitem{}
		Gehrels, N., Ramirez-Ruiz, E., \& Fox, D. B. (2009). 
		\textsl{Gamma-Ray Bursts in the Swift Era.} Annual Review of Astronomy and Astrophysics. \textbf{47}, 567--617.
		\bibitem{}
		Handl, J., Knowles, K. \& Kell, D. (2005). \textsl{Computational cluster validation in post-genomic data analysis}. Bioinformatics. \textbf{21}, 3201--3212.
		\bibitem{}
		Hakkila, J., Giblin, T. W., Roiger, R. J., Haglin, D. J., Paciesas, W. S., Meegan,
		C. A. (2003). \textsl{How Sample Completeness Affects Gamma-Ray Burst Classification.} The Astrophysical Journal. \textbf{582}, 320--329.
		\bibitem{}
		Hakkila, J., Haglin, D. J., Pendleton, G. N., Mallozzi, R. S., Meegan, C. A. \&
		Roiger, R. J. (2000). \textsl{Gamma-ray burst class properties}. The Astrophysical Journal. \textbf{538}, 165--180.
		\bibitem{}
		Hartigan, J. A. (1975). \textsl{Clustering Algorithms}. John Wiley \& Sons, New York, USA.
		\bibitem{}
		Hartigan, J. A. and Wong, M. A. (1979). \textsl{A K-means clustering algorithm}.
		Applied Statistics. \textbf{28}, 100--108.
		\bibitem{}
		Hogg, R. V., Mckean, J. W. and Craig, A. T. (2019). \textsl{Introduction
			to Mathematical Statistics}. Pearson Education, Boston.
		\bibitem{}
		Horv\'{a}th , I. (2002). \textsl{A further study of the BATSE Gamma-Ray Burst duration
			distribution}, Astronomy \& Astrophysics, \textbf{392}, 791–793.
		\bibitem{}
		Jain, A. K. , Murty, M. N. and Flynn, P. J. (1999). \textsl{Data clustering: a review}. ACM
		Computing Surveys. \textbf{31}, 264--323.
		\bibitem{}
		Jure\v{c}kov\'{a}, J. and Kalina, J. (2012). \textsl{Nonparametric multivariate
			rank tests and their unbiasedness.} Bernoulli, \textbf{18}, 229–-251.
		\bibitem{}
		Kaufman, L. and Rousseeuw, P. J. (2005). \textsl{Finding Groups in Data: An Introduction to Cluster Analysis.} John Wiley and Sons, New Jersey.
		\bibitem{}
		King, A., Olsson, E., \& Davies, M. B. (2007). \textsl{A new type of long gamma-ray burst.} Monthly Notices of the Royal Astronomical Society. \textbf{374}, L34.
		\bibitem{}
		Kost, J. T. and McDermott, M. P. (2002). \textsl{Combining dependent p-values.} Statistics \& Probability Letters, \textbf{60}, 183–-190.
		\bibitem{}
		Kouveliotou, C., Meegan, C. A., Fishman, G. J., Bhat, N. P., Briggs, M. S., Koshut, T. M., Paciesas, W. S., \& Pendleton, G. N. (1993). \textsl{Identification of two classes of gamma-ray bursts.} The Astrophysical Journal. \textbf{413}, L101.
		\bibitem{}
		Kulkarni, S., and Desai, S. (2017). \textsl{Classification of gamma-ray burst durations using robust model-comparison techniques}.  Astrophysics and Space Science, \textbf{362}, Article no. 70.
		{
			\bibitem{}
			Lamb, G. P., Tanvir, N. R., Levan, A. J., de Ugarte Postigo,  A., Kawaguchi, K., Corsi, A., Evans, P. A., Gompertz, B., Malesani, D. B., Page, K. L., Wiersema, K., Rosswog, S., Shibata, M., Tanaka, M., van der Horst, A. J., Cano, Z., Fynbo, J. P. U., Fruchter, A. S., Greiner, J., Heintz, K. E.,  Higgins, A., Hjorth, J., Izzo, L., Jakobsson, P., Kann, D. A., O'Brien, P. T., Perley, D. A., Pian, E., Pugliese, G., Starling, R. L. C., Thöne, C. C. , Watson, D., Wijers, R. A. M. J., and Xu, D. (2019). \textsl{Short GRB 160821B: A Reverse Shock, a Refreshed Shock, and a Well-sampled Kilonova}. The Astrophysical Journal, \textbf{883}, Article Number: 48, Pages: 12.
			\bibitem{}
			Levan, A., Crowther, P., de Grijs, R., Langer, N., Xu, D., Yoon, S.--C. (2016). \textsl{Gamma-Ray Burst Progenitors.}
			Space Science Reviews. \textbf{202}, 33--78.
			\bibitem{}
			Luo, J.-W., Wang, F.-F., Zhu-Ge, J.-M., Li, Y., Zou, Y.-C., Zhang, B. (2023). \textsl{Identifying the Physical Origin of Gamma-Ray Bursts with Supervised Machine Learning}, The Astrophysical Journal, \textbf{959},
			Article No. 44, Pages: 17.}
		\bibitem{}
		Mazets, E. P., Golenetskii, S. V., Ilyinskii, V. N., Panov, V. N., Aptekar, R. L., Guryan, Yu. A., Proskura, M. P., Sokolov, I. A., Sokolova, Z. Ya., Kharitonova, T. V., Dyatchkov, A. V., \& Khavenson, N. G (1981). \textsl{Catalog of cosmic gamma-ray bursts from the KONUS experiment data.} Astrophysics and Space Science. \textbf{80}, 119--143.
		\bibitem{}
		Matioli, L. C., Santos,  S. R., Kleina,  M. \& Leite, E. A. (2018). \textsl{A new algorithm for clustering based on kernel density estimation}. Journal of Applied Statistics. \textbf{45}, 347--366.
		\bibitem{}
		McLachlan, G. and Peel, D. (2000). \textsl{Finite Mixture Models}. John Wiley and Sons, New York.
		\bibitem{}
		Mehta, N. and Iyyani, S. (2024). \textsl{Exploring Gamma-Ray Burst Diversity: Clustering Analysis of the Emission Characteristics of Fermi- and BATSE-detected Gamma-Ray Bursts}. The Astrophysical Journal, \textbf{969}, Article id: 88, 12 pages.
		{
			\bibitem{}
			Melandri, A., Malesani, D. B., Izzo, L., Japelj, J., Vergani, S. D., Schady, P., Sagués Carracedo, A., de Ugarte Postigo, A., Anderson, J. P., Barbarino, C., Bolmer, J., Breeveld, A., Calissendorff, P., Campana, S., Cano, Z., Carini, R., Covino, S., D’Avanzo, P., D’Elia, V., della Valle, M., De Pasquale, M., Fynbo, J. P. U., Gromadzki, M., Hammer, F., Hartmann, D. H., Heintz, K. E., Inserra, C., Jakobsson, P., Kann, D. A., Kotilainen, J., Maguire, K., Masetti, N., Nicholl, M., Olivares E., F., Pugliese, G., Rossi, A., Salvaterra, R., Sollerman, J., Stone, M. B., Tagliaferri, G., Tomasella, L., Thöne, C. C., Xu, D., Young, D. R. (2019). \textsl{GRB 171010A/SN 2017htp: a GRB-SN at z = 0.33}, Monthly Notices of the Royal Astronomical Society, \textbf{490}, 5366–5374.
			\bibitem{}
			Minaev, P. Y. and Pozanenko, A. S. (2020). \textsl{The $E_{p,i}–E_{iso}$ correlation: type I gamma-ray bursts and the new
				classification method}. Monthly Notices of the Royal Astronomical Society, \textbf{492}, 1919–1936.}
		\bibitem{}
		Modak, S. (2019). \textsl{Uncovering astrophysical phenomena related to galaxies and other objects through statistical analysis.} Doctoral Thesis, University of Calcutta, URL: http://hdl.handle.net/10603/314773 
		\bibitem{}
		Modak, S. (2021). \textsl{Distinction of groups of gamma-ray bursts in the BATSE catalog through fuzzy clustering}. Astronomy and Computing. \textbf{34}, Article id 100441, 1--7.
		\bibitem{}
		Modak, S. (2022). \textsl{A new nonparametric interpoint distance-based measure for assessment of clustering}. Journal of Statistical Computation and Simulation. \textbf{9},	1062--1077.
		\bibitem{}
		Modak, S. (2023a). \textsl{Pointwise norm-based clustering of data in arbitrary dimensional space}. Communications in Statistics - Case Studies, Data Analysis and Applications, \textbf{9}, 121–134.
		\bibitem{}
		Modak, S. (2023b). \textsl{A new measure for assessment of clustering based on kernel	density estimation}. Communications in Statistics -- Theory and Methods, \textbf{52}, 5942-5951.
		\bibitem{}
		Modak, S. (2023c), \textsl{Validity index for clustered data in non-negative space}, Calcutta Statistical Association Bulletin, \textbf{75}, 60–71. 
		\bibitem{}
		Modak, S. (2023d), \textsl{Statistical Methods for Astronomical Data Analysis authored by Asis Kumar Chattopadhyay \& Tanuka Chattopadhyay}, Australian \& New Zealand Journal of Statistics, \textbf{65}, 394–395. 
		\bibitem{}
		Modak, S. (2024a). \textsl{Book Review: Finding Groups in Data: An Introduction to Cluster Analysis, Leonard Kaufman \& Peter J. Rousseeuw}, Journal of Applied Statistics, \textbf{51}, 1618--1620.
		\bibitem{}
		Modak, S. (2024b), \textsl{Evaluation of the number of clusters in a data set using p-values from multiple tests of hypotheses}, Communications in Statistics - Theory and Methods, \textbf{53}, 8878-8889.
		\bibitem{}
		Modak, S. (2024c). \textsl{A new interpoint distance-based clustering algorithm using kernel density estimation}, Communications in Statistics - Simulation and Computation, \textbf{53}, 5323-5341.
		\bibitem{}
		Modak, S. (2024d). \textsl{Determination of the number of clusters through logistic regression analysis}. Journal of Applied Statistics, \textbf{51}, 2344-2363.
		\bibitem{}
		Modak, S. (2024e). \textsl{A New Clustering Accuracy Measure Based on Relative Distances and its Cross-Validation Using Dirichlet Distribution}, Journal of Statistical Theory and Practice. \textbf{18}, Article no. 43, 14 Pages.
		\bibitem{}
		Modak, S. \& Bandyopadhyay, U. (2019). \textsl{A new nonparametric test for two sample multivariate location problem with application to astronomy}. Journal of Statistical Theory and Applications. \textbf{18}, 136--146.
		\bibitem{}
		Modak, S., Chattopadhyay, T. \& Chattopadhyay, A. K. (2017).
		\textsl{Two phase formation of massive elliptical galaxies: study through cross-correlation including spatial effect}, Astrophysics and Space Science. \textbf{362}, Article id: 206, pages 1--10.
		\bibitem{}
		Modak, S., Chattopadhyay, A. K. \& Chattopadhyay, T. (2018). \textsl{Clustering of gamma-ray bursts through kernel principal component analysis}. Communications in Statistics -- Simulation and Computation. \textbf{47}, 1088--1102.
		\bibitem{}
		Modak, S., Chattopadhyay, T. \& Chattopadhyay, A. K. (2020). \textsl{Unsupervised classification of eclipsing binary light curves through k-medoids
			clustering}. Journal of Applied Statistics. \textbf{47}, 376--392.
		\bibitem{}
		Modak, S., Chattopadhyay, T. \& Chattopadhyay, A. K. (2022). \textsl{Clustering of eclipsing binary light curves through functional principal component analysis}. Astrophysics and Space Science. \textbf{ 367}, Article id: 19, pages 1--10.
		\bibitem{}
		Mukherjee, S., Feigelson, E. D., Babu, G. J., Murtagh, F., Fraley, C. \& Raftery, A.
		(1998). \textsl{Three types of gamma-ray bursts}. The Astrophysical Journal. \textbf{508}, 314--327.
		\bibitem{}
		Nakar, E. (2007). \textsl{Short-hard gamma-ray bursts}. Physics Reports. \textbf{442}, 166--236.
		Narayana Bhat, P., Meegan, C. A., von Kienlin, A., et al. (2016). \textsl{The third Fermi GBM gamma-ray burst catalog: The first six years}. Astrophysics and Space Science. \textbf{223}, Article id: 28, 18 pages. 
		\bibitem{}
		Norris, J. P., Cline, T. L., Desai, U. D., \& Teegarden, B. J. (1984). \textsl{Frequency of fast, narrow $\gamma$-ray bursts.} Nature. \textbf{308},  434--435.
		\bibitem{}
		Ohmori, N., Yamaoka, K., Ohno, M., et al. (2016), \textsl{Suzaku Wide-band All-sky Monitor
			measurements of duration distributions of
			gamma-ray bursts}, Publications of the Astronomical Society of Japan, \textbf{68}, S30, 11 pages
		{\bibitem{}
			Paczy\'{n}ski, B., (1998). \textsl{Are gamma-ray bursts in star-forming regions$?$}.  The Astrophysical Journal. \textbf{494}, L45.}
		\bibitem{}
		Pakhiraa, M. K., Bandyopadhyay, S. and Maulik, U. (2004). \textsl{Validity index for crisp and fuzzy clusters}. Pattern Recognition. \textbf{37}, 487--501.
		\bibitem{}
		Poole, W., Gibbs, D. L., Shmulevich,. I., Bernard, B., Knijnenburg, T. A. (2016). \textsl{Combining dependent P-values with an
			empirical adaptation of Brown’s method.} Bioinformatics, \textbf{32}, i430–-i436.
		\bibitem{}
		Rajaniemi, H. J. \& M\"{a}h\"{o}nen, P. (2002). \textsl{Classifying Gamma-Ray Bursts using Self-organizing Maps.} The Astrophysical Journal. \textbf{566}, 202--209.
		\bibitem{}
		\v{R}\'{\i}pa, J.,  M\'{e}sz\'{a}ros, A., Veres, P., \& Park, I. H. (2012). \textsl{On the spectral lags and peak countsof the gamma-ray bursts detected by the RHESSI satellite}. The Astrophysical Journal, \textbf{756}, Article No. 44, 13 pages.
		\bibitem{}
		Ripley, B. D. (1996). \textsl{Pattern recognition and neural networks}. Cambridge University Press, Cambridge.
		\bibitem{}
		Rousseeuw, P. J. (1987). \textsl{Silhouettes: A graphical aid to the interpretation and validation of cluster analysis.} Journal of Computational and Applied Mathematics. \textbf{20}, 53--65. 
		\bibitem{}
		Silva, L. E. Brito Da, Melton, N. M. and Wunsch, D. C. (2020). \textsl{Incremental Cluster Validity Indices for Online Learning of Hard Partitions: Extensions and Comparative Study}. Institute of Electrical and Electronics Engineers, \textbf{8}, 22025--22047.
		\bibitem{}
		Scrucca, L., Fop, M., Murphy, T. B. and Raftery, A. E. (2016). \textsl{mclust 5: Clustering, Classification and Density Estimation Using Gaussian Finite Mixture Models}. The R Journal, \textbf{8}, 289–317.
		\bibitem{}
		Tibshirani, R., Walther, G. and Hastie, T. (2001). \textsl{Estimating of the number of clusters in data set via the gap statistic}, Journal of Royal Statistical Society, Series B, \textbf{63}, 411-423.
		\bibitem{}
		Tarnopolski, M. (2015). \textsl{On the limit between short and long GRBs}, Astrophysics and Space Science. \textbf{359:}, 20.
		\bibitem{}
		Tarnopolski, M. (2016a). \textsl{Analysis of gamma-ray burst duration distribution using mixtures of skewed distributions}, Monthly Notices of the Royal Astronomical Society. \textbf{458}, 2024–2031.
		\bibitem{}
		Tarnopolski, M. (2016b). \textsl{Analysis of the observed and intrinsic durations of Swift/BAT
			gamma-ray bursts}, New Astronomy, \textbf{46}, 54–59.
		\bibitem{}
		Tarnopolski, M. (2019). \textsl{Analysis of the Duration–Hardness Ratio Plane of Gamma-Ray Bursts Using Skewed
			Distributions}, The Astrophysical Journal. \textbf{870}, Article id: 105, 9 pages.
		\bibitem{}
		Tarnopolski, M. (2022). \textsl{Graph-based clustering of gamma-ray bursts}. Astronomy \& Astrophysics, \textbf{657}, Article No. A13, 8 pages.
		\bibitem{}
		T\'{o}th, B. G., R\'{a}cz, I. I. \& Horv\'{a}th, I. (2019). \textsl{Gaussian-mixture-model-based cluster analysis of gamma-ray bursts in
			the BATSE catalog}. Monthly Notices of the Royal Astronomical Society. \textbf{486}, 4823--4828.
		{
			\bibitem{}
			Troja, E., Castro-Tirado, A. J., González, J. B., Hu, Y., Ryan, G. S., Cenko, S. B., Ricci, R., Novara, G., Sánchez-Rámirez, R., Acosta-Pulido, J. A., Ackley, K. D.,  García, M. D. C., Eikenberry, S. S., Guziy, S., Jeong, S., Lien, A. Y., Márquez,  I., Pandey, S. B., Park, I. H., Sakamoto, T., Tello, J. C., Sokolov, I. V., Sokolov,  V. V., Tiengo, A., Valeev, A. F., Zhang, B. B., Veilleux, S. (2019). \textsl{The afterglow and kilonova of the short GRB 160821B.} Monthly Notices of the Royal Astronomical Society, \textbf{489}, 2104–2116.}
		\bibitem{}
		Yang, E. B., Zhang, Z. B., \& Jiang, X. X. (2016). \textsl{Two dimensional classification of the Swift/BAT GRBs}.  Astrophysics and Space Science, \textbf{361}, Article id: 257
		{
			\bibitem{}
			Usov, V. V. (1992). \textsl{Millisecond pulsars with extremely strong magnetic fields as a cosmological source of $\gamma$-ray bursts}. Nature. \textbf{357}, 472--474.
			\bibitem{}
			Wang, H., Zhang, F.-W., Wang,  Y.-Z., Shen, Z.-Q., Liang, Y.-F., Li,  X., Liao, N.-H., Jin, Z.-P., Yuan,  Q., Zou, Y.-C., Fan, Y.-Z., and Wei, D.-M. (2017), \textsl{The GW170817/GRB 170817A/AT 2017gfo Association: Some Implications for Physics and Astrophysics}. The Astrophysical Journal Letters, \textbf{851}, Article id: L18, 7 Pages.}
		\bibitem{}
		Woosley, S. E. \& Bloom, J. S. (2006). \textsl{The Supernova Gamma-Ray Burst Connection}. Annual Review of Astronomy \& Astrophysics. \textbf{44}, 507--556.
		\bibitem{}
		Zhang, F.--W., Shao, L., Yan J.--Z., \& Wei, D.--M. (2012). \textsl{Revisiting the Long/Soft-Short/Hard Classification of Gamma-Ray Bursts in the Fermi Era
			.} The Astrophysical Journal. \textbf{750}, 88.
		{
			\bibitem{}
			Zhu, S.-Y., Sun, W.-P., Ma, D.-L. and Zhang, F.-W. (2024). \textsl{Classification of Fermi gamma-ray bursts based on machine learning.}Monthly Notices of the Royal Astronomical Society. \textbf{532}, 1434–1443. }
		\bibitem{}
		Zitouni, H., Guessoum, N., AlQassimi, K. M., \& Alaryani, O. (2018). \textsl{Distributions of pseudo-redshifts and durations (observed and intrinsic) of Fermi GRBs}. Astronomy \& Astrophysics, 
		\textbf{363}, Article No. 223.
	\end{thebibliography}
\end{document}